\documentclass[pra,twocolumn,aps,floatfix,longbibliography,superscriptaddress]{revtex4-1}
\UseRawInputEncoding
\usepackage{comment}
\usepackage[utf8]{inputenc}
\usepackage{mathrsfs}
\usepackage{physics}
\usepackage{tikz}
\usepackage[caption=false]{subfig}
\usepackage{wrapfig}
\usepackage[colorlinks=true,linkcolor=blue,citecolor=blue,urlcolor=blue]{hyperref}
\usepackage{amsmath,amssymb,amsfonts,amsthm}
\usepackage{soul}

\usepackage{graphicx}
\usepackage{ragged2e}
\usepackage{float}
\usepackage{lipsum}

\newcommand{\kk}{\mathbf{k}}
\newcommand{\xx}{\mathbf{x}}

\usepackage[normalem]{ulem}

\usepackage{xcolor}

\begin{document}
\date{\today}
\author{Nikolaos K. Kollas}
\email{kollas@upatras.gr}
\author{Dimitris Moustos}
\email{dmoustos@upatras.gr}
\affiliation{Division of Theoretical and Mathematical Physics, Astronomy and Astrophysics, Department of Physics, University of Patras, 26504,  Patras, Greece}
\author{Miguel R. Mu{\~n}oz}
\affiliation{Department of Physics and Astronomy,  University of Sheffield, Sheffield, S3 7RH,United Kingdom}
\title{Cohering and decohering power of massive scalar fields under instantaneous interactions}
\begin{abstract}
Employing a non-perturbative approach based on an instantaneous interaction between a two-level Unruh-DeWitt detector and a massive scalar field, we investigate the ability of the field to generate or destroy coherence in the detector by deriving the cohering and decohering power of the induced quantum evolution channel. For a field in a coherent state a previously unnoticed effect is reported whereby the amount of coherence that the field generates displays a revival pattern with respect to the size of the detector. It is demonstrated that by including mass in a thermal field the set of maximally coherent states of the detector decoheres less compared to a zero mass. In both of the examples mentioned, by making a suitable choice of detector radius, field energy and coupling strength it is possible to infer the mass of the field by either measuring the coherence present in the detector in the case of an interaction with a coherent field or the corresponding decoherence of a maximally coherent state in the case of a thermal field. In view of recent advances in the study of Proca metamaterials, these results suggest the possibility of utilising the theory of massive electromagnetism for the construction of novel applications for use in quantum technologies.
\end{abstract}
\maketitle
%--------------------
%
%--------------------
\section{Introduction}
Coherent systems, defined as existing in a superposition of different states, form the backbone of the second quantum revolution brought about by the advent of quantum information science and technology \cite{nielsen_chuang_2010,doi:10.1098/rsta.2003.1227}. Formalized in terms of a quantum resource theory \cite{aberg2006quantifying,PhysRevLett.113.140401,PhysRevLett.116.120404,RevModPhys.89.041003,PhysRevLett.119.230401,https://doi.org/10.1002/qute.202100040} the equivalence between coherence and entanglement (the fuel behind applications such as quantum dense coding \cite{PhysRevLett.69.2881}, unhackable cryptography \cite{PhysRevLett.67.661} and teleportation \cite{PhysRevLett.70.1895}) was recognized early on \cite{PhysRevLett.115.020403,PhysRevLett.117.020402,PhysRevLett.128.160402}. Recently, the role of coherence and its depletion during the execution of quantum algorithms has received increasing attention \cite{PhysRevA.93.012111,arxiv.2205.13610,arxiv.2203.10632,arxiv.2112.10867,Pan_2022,PhysRevA.95.032307,e21030260}. Coherence plays a part in other physical contexts as well, such as in quantum metrology \cite{PhysRevA.94.052324,Giorda_2017}, thermodynamics \cite{Lostaglio2015,PhysRevLett.115.210403,PhysRevX.5.021001,Narasimhachar2015,Korzekwa_2016} and even possibly in biological processes \cite{Lloyd_2011,doi:10.1080/00405000.2013.829687}. Because of its usefulness as a resource, it is of particular interest to study the conditions under which coherence can be extracted or generated from other systems \cite{PhysRevLett.113.150402,PhysRevA.101.042325,Swelling,NK:DM}, as well as devise methods for its protection \cite{PhysRevA.99.022107,PhysRevA.104.052405,Miller_2022} against the decohering effects of the environment \cite{PhysRevA.51.992,buchleitner2002coherent,schlosshauer}.

In this report we will examine the ability of a massive quantum field to generate or destroy coherence in a two-level Unruh-DeWitt (UDW) detector under an instantaneous interaction \cite{Simidzija,https://doi.org/10.48550/arxiv.2002.01994,PhysRevD.101.036014,PhysRevD.105.065011,PhysRevD.104.125017, PhysRevD.105.085011,https://doi.org/10.48550/arxiv.2204.02983} (for a study of the coherence present in the field under a different context see \cite{Huang2018,Du2017,Wu_2021,PhysRevA.105.052403}). To accomplish this we will determine the cohering and decohering power of the quantum evolution channel induced by the action of the field on the detector \cite{PhysRevA.92.032331,BuXiong,10.5555/3179439.3179441,BU20171670,PhysRevA.105.L060401}. Compared to other approaches that study coherence in a relativistic setting in a perturbative manner \cite{https://doi.org/10.48550/arxiv.2111.01358,Feng_2022,PhysRevA.93.062105,Swelling,NK:DM,Huang2022,coh:axions}, an instantaneous interaction permits an exact solution of the final state of the detector for arbitrary coupling strengths. This provides the opportunity for a better understanding of the effects that different parameters such as the size of the detector, the energy of the field or its temperature have in the creation and destruction of coherence, free from any need for use of approximations. An example is given in Section \ref{SecIV}, where it is observed that for specific values of the detector's radius, the amount of coherence generated by a coherent field vanishes, an effect which in a perturbative treatment would have otherwise remained unnoticed.

The reasons for considering a scalar field with mass will become apparent in Section \ref{SecV}, where the decohering power of a thermal field with inverse temperature $\beta$ is presented. The ability of the field to preserve part of the coherence present in a maximally coherent state of the detector is enhanced for increasing values of its mass. This observation is in line with similar perturbative results about the coherent behaviour of an atom immersed in a massive field \cite{Huang2022} and the advantages of mass in entanglement harvesting  \cite{Harvest:mass,JHEPentmass,https://doi.org/10.48550/arxiv.2206.06381} and sensing \cite{Quantaccel,QFItherm}. Since decoherence is currently a major hurdle in practical uses of quantum computation such results may be of interest and could perhaps be leveraged with the use of massive electromagnetic fields in Proca metamaterials \cite{Procameta}. 

In \cite{coh:axions} the authors considered the possibility of using the coherence of the detector as a means of probing the mass of axion dark matter. We show how, under a suitable choice of parameters, it is similarly possible to infer the mass of a scalar field by either measuring the cohering power of a coherent or the decohering power of a thermal state of the field. In this case changes in coherence are easier to detect since they are orders of magnitude larger than what is possible in a weak coupling treatment.

We begin by giving a short introduction in Section \ref{SecII} to the resource theory of quantum coherence and the UDW detector model in Section \ref{SecII}. As is common practice, throughout the manuscript we employ a natural system of units in which $\hbar=c=k_B=1$.
%---------------------------------------------------------
% 
%---------------------------------------------------------
\section{Cohering and decohering power of quantum channels}\label{SecII}
Coherence, i.e. the degree of superposition of a quantum system \cite{aberg2006quantifying,RevModPhys.89.041003,PhysRevLett.119.230401}, is dependent on the choice of basis of the underlying Hilbert space in which we decide to express the state $\rho$ of the system.  For a state of the form
\begin{equation}
    \rho=\sum_{i,j}\rho_{ij}\ketbra{i}{j},
\end{equation}
where $\{\ket{i}\}_{i=0}^{d-1}$ is a finite set of basis spanning the $d$-dimensional Hilbert space $\mathbb{C}^d$, we say that $\rho$ represents a \emph{coherent state} with respect to this basis, if there exists at least one pair of indices $i\neq j$ such that $\rho_{ij}\neq 0$. A system which is \emph{incoherent} is represented by a diagonal matrix and satisfies
\begin{equation}
    \Delta(\rho) = \rho
\end{equation}
where 
\begin{equation}
    \Delta(\rho)=\sum_{i}\rho_{ii}\ketbra{i}
\end{equation}
denotes the \emph{dephasing operation} in the chosen basis.

The set of quantum operations acting on a state is similarly divided into those that can and those that cannot create coherence. The so called \emph{maximally incoherent operations} (MIO) are defined as those completely positive and trace preserving operations $\Phi$ that map the set of incoherent states $\mathcal{I}$ onto a subset of itself
\begin{equation}\label{MIO}
    \Phi(\mathcal{I})\subseteq\mathcal{I}.
\end{equation}
The ability of a quantum channel $\Phi$ to generate coherence out of incoherent states can be determined by calculating its \emph{cohering power}. \cite{PhysRevA.92.032331,BuXiong,10.5555/3179439.3179441,BU20171670,PhysRevA.105.L060401}. In order to define the latter it is necessary first to introduce the notion of a \emph{coherence measure}. This is a non-negative real valued function $C$ on the set of density matrices with the following properties:
\begin{enumerate}
    \item[i)] $C(\rho)\geq0$ with equality if and only if $\rho\in\mathcal{I}.$
    \item[ii)] $C(\Phi(\rho))\leq C(\rho)$ for every $\Phi$ $\in$ (MIO).
    \item[iii)] $C(\sum_ip_i\rho_i)\leq\sum_ip_iC(\rho_i)$.
\end{enumerate}
The first property requires the measure to be \emph{faithful} so that it can distinguish between coherent and incoherent states. The second property reflects the restrictions of the theory. Since by definition (MIO)'s cannot generate coherent out of incoherent states it makes sense to require the measure to be \emph{monotonic}, the amount of coherence in a state after the action of a (MIO) operation should therefore always be less than before. This property is what gives the theory the structure of a \emph{quantum resource} \cite{RevModPhys.91.025001}. The final property, which imposes \emph{convexity} on the measure, states that it is not possible to increase the average amount of coherence in a quantum ensemble $\{p_i,\rho_i\}$, where $p_i$ is the probability of obtaining state $\rho_i$, by simply mixing its elements.

Armed with a valid measure of coherence we are now able to define the cohering power of the channel as the maximum amount of coherence created when $\Phi$ acts on the set of incoherent states
\begin{equation}\label{continous coh power}
    \mathcal{C}(\Phi) = \max_{\rho\in\mathcal{I}}C(\Phi(\rho)).
\end{equation}
Because of convexity the maximum on the right hand side is actually reached by acting $\Phi$ on one of the basis states. This simplifies considerably the calculation since the required optimization is now performed over a discrete instead of a continuous set. In this case
\begin{equation}\label{cohering power}
    \mathcal{C}(\Phi) = \max_{\ket{i}}C(\Phi(\ketbra{i})).
\end{equation}

Another property of interest for a quantum channel is the amount of coherence that it destroys when it is applied on a \emph{maximally coherent state}, i.e. a uniform superposition, of the form
\begin{equation}\label{maximally coherent}
    \psi_d(\boldsymbol\theta)=\frac{1}{\sqrt{d}}\sum_{j=0}^{d-1}e^{i\theta_j}\ket{j}.
\end{equation}
Similar to Eq. (\ref{cohering power}) we define the \emph{decohering power} of the channel as the maximum possible difference in the amount of coherence before and after its action on the maximally coherent state
\begin{equation}
    \mathcal{D}(\Phi) = \max_{\boldsymbol\theta}[C(\psi_d(\boldsymbol\theta))-C(\Phi(\psi_d(\boldsymbol\theta)))].
\end{equation}

In what follows we will employ the oft-used $\ell_1$-norm of coherence as our measure. This is defined as the sum of the absolute values of the non-diagonal elements of the density matrix
\begin{equation}\label{l1-norm}
    C_{\ell_1}(\rho) = \sum_{i\neq j}\abs{\rho_{ij}}.
\end{equation}
For the set of maximally coherent states 
\begin{equation}
    C_{\ell_1}(\psi_d(\boldsymbol\theta))=d-1
\end{equation} 
so in this case
\begin{equation}\label{decohering power}
    \mathcal{D}_{\ell_1}(\Phi) = d-1 -\min_{\boldsymbol\theta}C_{\ell_1}(\Phi(\psi_d(\boldsymbol\theta))).
\end{equation}

%-----------------------------------------------------------------------------------------------------
%
%-----------------------------------------------------------------------------------------------------
\section{The Unruh-DeWitt detector model}\label{Sec III}
The UDW detector model is frequently employed as a means of studying the interaction between a two-level system (the detector) and a quantum field \cite{Unruh,DeWitt,birrell}. The interaction induces transitions between the detector's excited $\ket{e}$ and ground $\ket{g}$ states, which depend on the initial state of the field $\sigma_\phi$ as well as on the trajectory of the detector and the structure of the underlying spacetime. Coupling the \emph{monopole} operator of the detector
\begin{equation}\label{monopole}
    \hat{\mu}(t)=e^{i\Omega t}\ketbra{e}{g}+ e^{-i\Omega t}\ketbra{g}{e},
\end{equation}
to the field operator $\hat\varphi(t,\xx)$ evaluated at the detector's position $\xx$ at time $t$ defines the UDW interaction Hamiltonian
\begin{equation}\label{UDW:interaction}
    \hat{H}_{\text{int}}(t)=\chi(t)\hat{\mu}(t)\otimes\hat{\varphi}(t,\xx).
\end{equation}
where $\Omega$ is the energy gap between the detector's levels, and the real valued \emph{switching function} function $\chi(t)$ describes the strength of the interaction at each instant in time. For a scalar field with mass $m$ the field operator in flat Miknowski spacetime is equal to
\begin{equation}
    \hat\varphi(t,\xx)=\int\frac{d^3\kk}{\sqrt{(2\pi)^32\omega(\kk)}}\left(\hat{a}_{\kk}e^{i(\kk\cdot\xx-\omega(\kk)t)}+\text{H.c.}\right)
\end{equation}
where $\hat a_{\kk}$ and $\hat a_{\kk}^\dag$ denote the annihilation and creation operators respectively, of a field mode with momentum $\kk$ and energy $\omega(\kk)=\sqrt{|\kk|^2+m^2}$, that satisfy the canonical commutation relations
\begin{equation}\label{commut:rel}
    [\hat{a}_{\kk},\hat{a}_{\kk'}]=[\hat{a}_{\kk}^\dag,\hat{a}_{\kk'}^\dag]=0,\quad
    [\hat{a}_{\kk},\hat{a}_{\kk'}^\dag]=\delta(\kk-\kk').
\end{equation}

The UDW Hamiltonian describes a point-like interaction in which the field interacts with the detector at a single point in space each time. It is possible to take into account the finite size of the detector by averaging over a region in a neighborhood of the detector's position. For a detector at rest at position $\xx$ \footnote{Equation (\ref{smeared_interaction}) can be easily extended in the case of a moving detector by making use of a Fermi-Walker coordinate system \cite{MTW}.}, Eq. (\ref{UDW:interaction}) is replaced by
\begin{equation}\label{smeared_interaction}
    \hat{H}_{\text{int}}(t)=\chi(t)\hat{\mu}(t)\otimes\int f(\xx-\xx ')\hat{\varphi}(t,\xx ')d^3\xx '.
\end{equation}
The real valued  \emph{smearing function} function $f(\xx)$ with dimensions $(length)^{-3}$ reflects the shape and size of the detector \cite{Schlicht_2004,Louko_2006,Wavepacket:det,CHLI2} with a mean effective radius $R$ equal to
\begin{equation}\label{radius}
    R=\int\abs{\xx}f(\xx)d^3\xx.
\end{equation}
By taking the pointlike limit $f(\xx)=\delta(\xx)$, (i.e., $R\to 0$), Eq. (\ref{UDW:interaction}) is immediately recovered. Setting
\begin{equation}\label{Fourier}
    F(\kk)=\int f(\xx)e^{i\kk\cdot\xx}d^3\xx, 
\end{equation}
for the Fourier transform of the smearing function,  we can rewrite Eq. (\ref{smeared_interaction}) as
\begin{equation}\label{interaction}
    \hat{H}_{\text{int}}(t)=\chi(t)\hat{\mu}(t)\otimes\hat{\varphi}_f(t,\xx)
\end{equation}
with a `smeared' field operator of the form
\begin{equation}\label{mod_smeared}
\hat{\varphi}_f(t,\xx)=\int\frac{d^3\kk}{\sqrt{(2\pi)^32\omega(\kk)}}\left(F(\kk)\hat a_{\kk}e^{i(\kk\cdot\xx-\omega(\kk)t)}+\text{H.c.}\right).
\end{equation}

\subsection{Evolution under an instantaneous interaction}
In order to obtain the final state of the detector after the interaction with the field has been switched off, we must first evolve the combined system of detector and field with the unitary operator $\hat{U}$ generated by the time integral of the interaction Hamiltonian
\begin{equation}\label{evolution}
    \hat{U}=\mathcal{T}\text{exp}\left(-i\int\limits_{-\infty}^{+\infty}\hat{H}_{\text{int}}(t)dt\right),
\end{equation}
where $\mathcal{T}$ denotes the time ordering operator. Tracing out the field degrees of freedom, induces a quantum evolution channel on the initial state $\rho$ of the detector defined by
\begin{equation}\label{Phi definition}
\Phi(\rho)=\tr_{\varphi}[\hat{U}(\rho\otimes\sigma_\varphi)\hat{U}^\dagger].
\end{equation}
Under a delta switching function centered around $t_0$,
\begin{equation}
    \chi(t)=\lambda\delta(t-t_0),
\end{equation}
with $\lambda$ a coupling constant with the same dimensions as length, it is possible to drop the time ordering in \eqref{evolution} \cite{Simidzija,https://doi.org/10.48550/arxiv.2002.01994,PhysRevD.101.036014,PhysRevD.105.065011,PhysRevD.104.125017, PhysRevD.105.085011,https://doi.org/10.48550/arxiv.2204.02983}. In this case
\begin{equation}
    \hat{U}=\exp[-i\lambda\hat\mu_0\otimes\hat\varphi_{f_0}],
\end{equation}
where $\hat\mu_0=\hat\mu(t_0)$ and $\hat\varphi_{f_0}=\hat\varphi_{f}(t_0,\xx)$. With a little bit of algebra it is easy to show that since $\hat\mu_0^2=I$
\begin{equation}\label{exact unitary}
    \hat{U}=\frac{I-\hat\mu_0}{2}\otimes\exp(i\lambda\hat\varphi_{f_0})+\frac{I+\hat\mu_0}{2}\otimes\exp(-i\lambda\hat\varphi_{f_0}).
\end{equation}

Inserting Eq. (\ref{exact unitary}) into Eq. (\ref{Phi definition}), we find that the action of the channel on the detector is equal to a convex combination of a bit flip channel \cite{nielsen_chuang_2010}
\begin{equation}
    B(\rho)=\frac{\rho+\hat\mu_0\rho\hat\mu_0}{2}
\end{equation}
and a unitary rotation,
\begin{equation}
    V = \sqrt{\frac{\abs{z}+\Re z}{2\abs{z}}}\hat{I}+\sqrt{\frac{\abs{z}-\Re z}{2\abs{z}}}\hat\mu_0
\end{equation}
\begin{equation}\label{final state}
    \Phi(\rho)=(1-\abs{z})B(\rho)+\abs{z}V\rho V^\dagger
\end{equation}
%\begin{equation}\label{final state}
%    \Phi(\rho)=\frac{1}{2}\left[(1+\Re z)\rho+i[\rho,\hat\mu_0]\Im z+(1-\Re z)\hat\mu_0\rho\hat\mu_0\right],
%\end{equation}
where
\begin{equation}\label{z}
    z=\tr_\varphi[e^{i2\lambda\hat\varphi_{f_0}}\sigma_\varphi].
\end{equation}
\subsection{Cohering and decohering power of scalar fields}
According to Eqs (\ref{cohering power}) and (\ref{l1-norm}), the $\ell_1$-cohering power of the channel induced by the UDW interaction of the detector with the massive field, is equal to the maximum amount of coherence obtained by acting $\Phi$ on either the ground or excited state. In both cases this amount is the same and equal to
\begin{equation}\label{Phi cohering power}
    \mathcal{C}_{\ell_1}(\Phi) = \abs{\langle\sin(2\lambda\hat\varphi_{f_0})\rangle},
\end{equation}
where $\langle\hat X\rangle=\tr_\varphi (\hat X\sigma_\varphi)$ denotes the expectation value of field operator $\hat{X}$. This is actually equal to the maximum amount of coherence that can be obtained by acting $\Phi$ on the whole set of states (for more details consult Appendix  \ref{first appendix}). 

To obtain the $\ell_1$-decohering power requires a little more effort. Replacing the maximally coherent state
\begin{equation}
    \psi_2(\theta) = \frac{1}{\sqrt{2}}(\ket{g}+e^{i\theta}\ket{e})
\end{equation}
in Eq. (\ref{final state}) we see that the coherence of the final state of the detector is equal to
\begin{multline}\label{remaining coh}
    C_{\ell_1}(\Phi(\psi_2(\theta))) =\\ \sqrt{\cos^2\left(\frac{\theta-\Omega t_0}{2}\right)+(\Re z)^2\sin^2\left(\frac{\theta-\Omega t_0}{2}\right)}.
\end{multline}
Note that for a maximally coherent state with $\theta=\Omega t_0$ the amount of coherence before and after the interaction has taken place is frozen \cite{PhysRevLett.114.210401}. For this choice of phase the state is a fixed point of the evolution channel. This observation holds in general and is independent of details such as the mass of the field, its initial state or the size of the detector.

It is straightforward to show that the minimum in Eq. (\ref{remaining coh}) is obtained by setting $\theta=\pi+\Omega t_0$. With the help of Eq. (\ref{decohering power}) we therefore find that the $\ell_1$-decohering power of the field induced channel is equal to
\begin{equation}\label{Phi decohering power}
    \mathcal{D}_{\ell_1}(\Phi)=1-\abs{\langle\cos(2\lambda\hat\varphi_{f_0})\rangle}.
\end{equation}

We now proceed to study the cohering and decohering power of a field in a coherent and a thermal state respectively.

\section{Cohering power of coherent scalar fields}\label{SecIV}
\begin{figure*}\begin{minipage}{\textwidth}
\subfloat[${\lambda}=0.2\lambda_C$]{\includegraphics[width=0.45\textwidth]{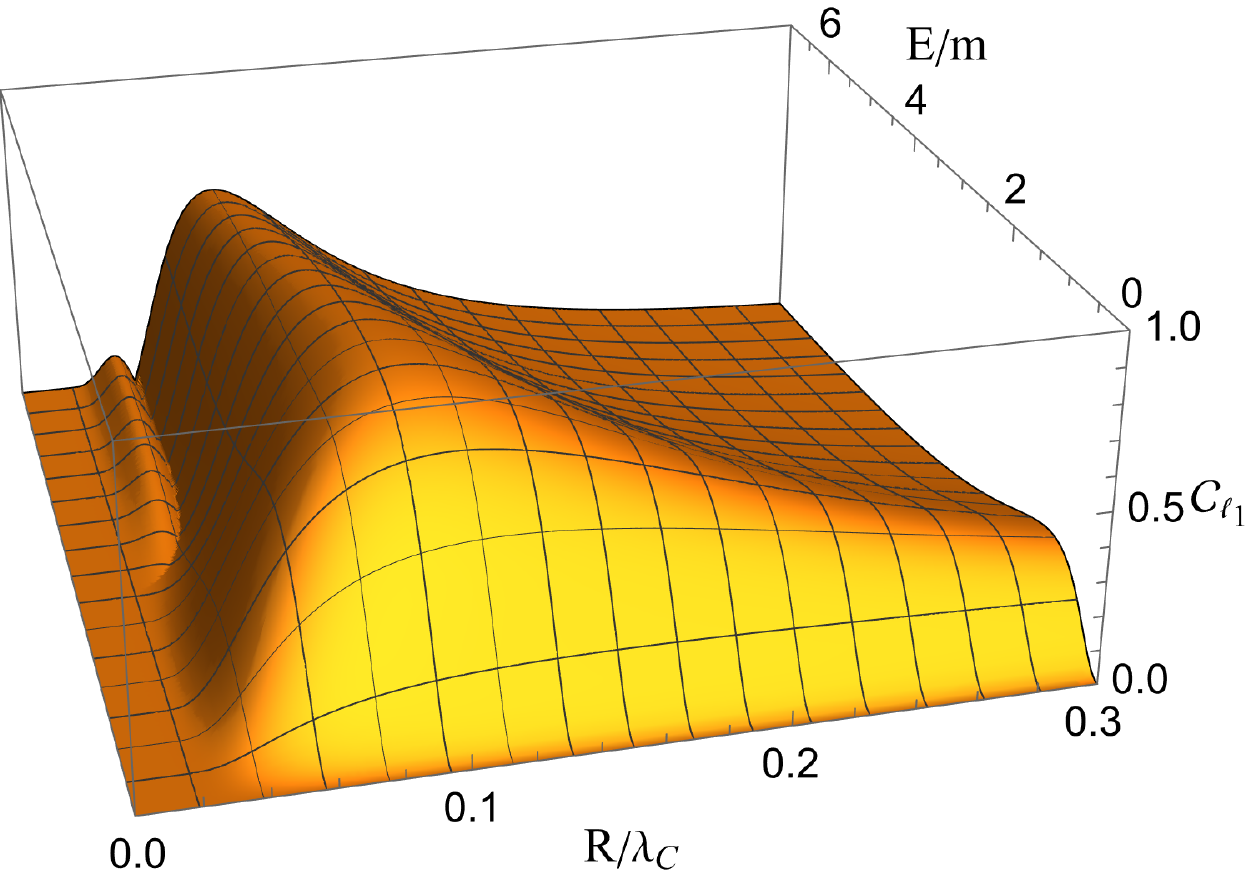}}\hspace{0.5cm}
\subfloat[${\lambda}=\lambda_C$]{\includegraphics[width=0.45\textwidth]{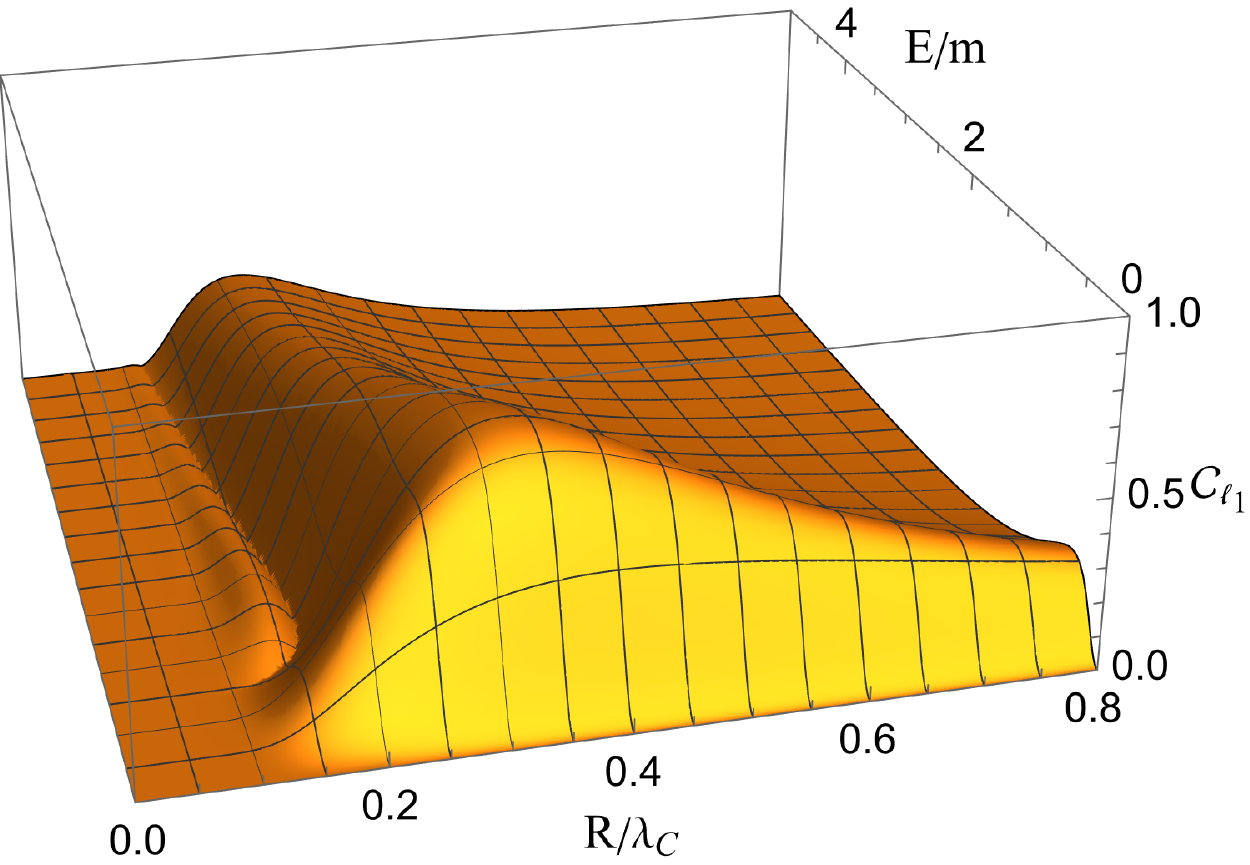}}\\
\subfloat[${\lambda}=5\lambda_C$]{\includegraphics[width=0.45\textwidth]{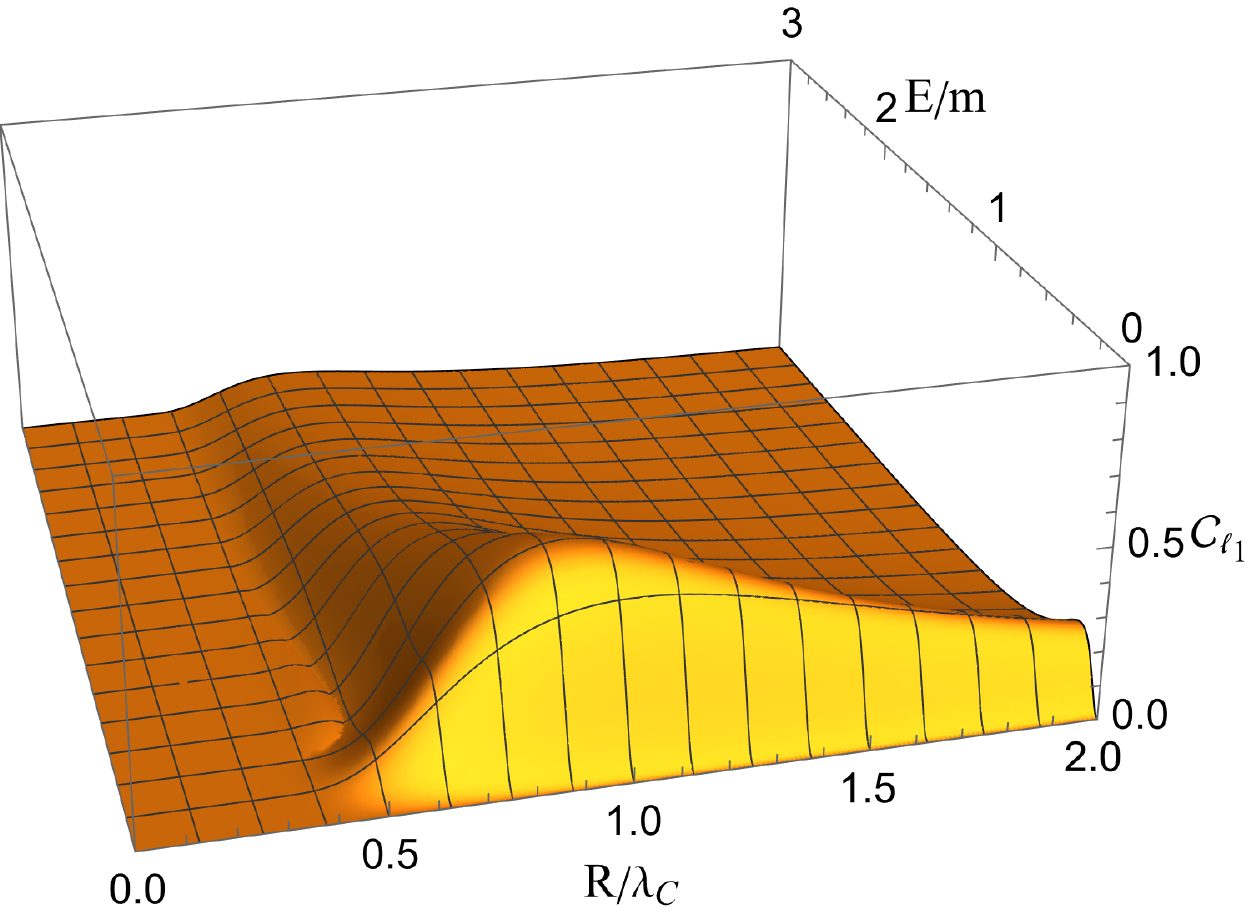}}\hspace{0.5cm}
\subfloat[${\lambda}=\lambda_C$]{\includegraphics[scale=0.45]{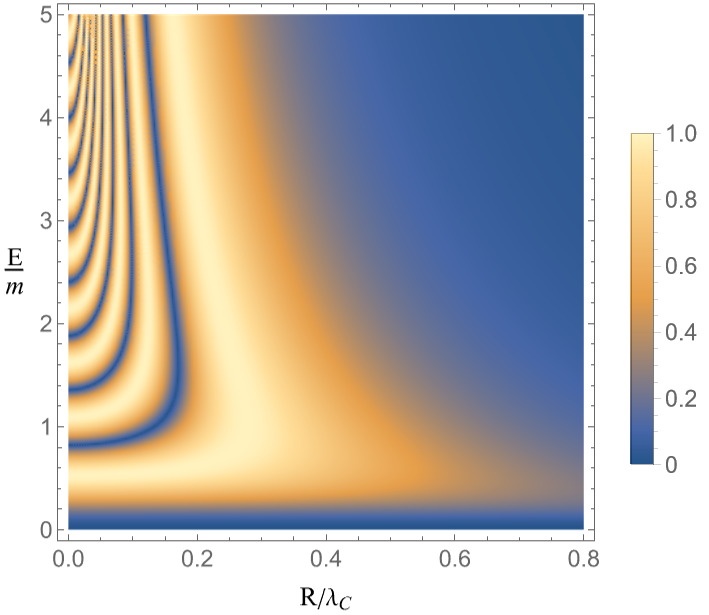}}
\end{minipage}
\caption{(a)-(c): $\ell_1$-cohering power of a quantum evolution channel acting on a static Unruh-DeWitt detector as a result of its interaction with a massive scalar field in a coherent state under an instantaneous interaction, as a function of the energy of the field in field mass units, $E/m$, and the effective radius of the detector in Compton wavelength units $R/\lambda_C$ for three different values of the coupling constant between detector and field. (d): Oscillatory behavior of $\ell_1$-cohering power for $\lambda=\lambda_C$.}
\label{fig:1}
\end{figure*}

A coherent state $\ket{a}$ of the field is equivalent to a complex valued \emph{coherent amplitude distribution $a(\kk)$} such that the action of the annihilation operator $\hat{a}_\kk$ on the state is equal to \cite{PhysRev.130.2529,Simidzija,PhysRevD.96.025020}
\begin{equation}
    \hat{a}_{\kk}\ket{a}=a(\kk)\ket{a}.
\end{equation} 
Let us now decompose the field into parts
\begin{equation}\label{decomposed field}
    \hat\varphi_{f_0} = \hat{\mathrm{a}} + \hat{\mathrm{a}}^\dagger
\end{equation}
each containing only annihilation or creation operators
\begin{subequations}
\begin{equation}\label{phi-a}
    \hat{\mathrm{a}} = \int\frac{d^3\kk}{\sqrt{(2\pi)^32\omega(\kk)}}F(\kk)\hat a_{\kk}e^{i(\kk\cdot\xx-\omega(\kk)t_0)},
\end{equation}
\begin{equation}
    \hat{\mathrm{a}}^\dagger = \int\frac{d^3\kk}{\sqrt{(2\pi)^32\omega(\kk)}}F^*(\kk)\hat a^\dagger_{\kk}e^{-i(\kk\cdot\xx-\omega(\kk)t_0)}.
\end{equation}
\end{subequations}
By employing the Baker-Campbell-Hausdorff formula
\begin{equation}
    e^{\hat{X}+\hat{Y}}=e^{\hat{X}}e^{\hat{Y}}e^{-\frac{1}{2}[\hat{X},\hat{Y}]},
\end{equation}
which holds true when both $[\hat{X},[\hat{X},\hat{Y}]]=0$ and $[\hat{Y},[\hat{X},\hat{Y}]]=0$, it can be shown that
\begin{align}\label{zBCH}
    \langle e^{i2\lambda\hat\varphi_{f_0}}\rangle_a&=e^{-2\lambda^2[\hat{\mathrm{a}},\hat{\mathrm{a}}^\dagger]}\langle e^{i2\lambda \hat{\mathrm{a}}^\dagger}e^{i2\lambda\hat{\mathrm{a}}}\rangle_a.\nonumber\\
    &=e^{-2\lambda^2[\hat{\mathrm{a}},\hat{\mathrm{a}}^\dagger]}e^{i4\lambda\Re \langle\hat{\mathrm{a}}\rangle_a}
\end{align}
where
\begin{equation}
    [\hat{\mathrm{a}},\hat{\mathrm{a}}^\dagger]=\frac{1}{(2\pi)^3}\int\frac{\abs{F(\kk)}^2}{2\omega(\kk)}d^3\kk,
\end{equation}
and the subscript in the expectation value of the field operator is included in order to indicate its dependence on the coherent amplitude distribution.

From Eq. (\ref{Phi cohering power}) it follows that the $\ell_1$-cohering power of a coherent scalar field is equal to
\begin{equation}\label{coher:exp}
    \mathcal{C}_{\ell_1}(\Phi)= e^{-2\lambda^2[\hat{\mathrm{a}},\hat{\mathrm{a}}^\dagger]}\abs{\sin({4\lambda\Re \langle\hat{\mathrm{a}}\rangle_a})}.
\end{equation}
Assuming a static detector with a Gaussian smearing function and a mean effective radius equal to $R$
\begin{equation}\label{smear:gauss}
f(\xx)=\frac{\exp[-\frac{4\abs{{\xx}}^2}{\pi R^2}]}{(\pi R/2)^3},
\end{equation}
with a corresponding Fourier transform of the form
\begin{equation}
    F(\kk)= \text{exp}\left[-\frac{\pi \abs{\kk}^2R^2}{16}\right],
\end{equation}
the commutator between $\hat{\mathrm{a}}$ and $\hat{\mathrm{a}}^\dagger$ will now dependson the radius and mass of the field and will be equal to
\begin{align}\label{mass-com}
    [\hat{\mathrm{a}},\hat{\mathrm{a}}^\dagger]&=\frac{1}{(4\pi^2)}\int_{0}^\infty\frac{k^2e^{-\frac{\pi k^2R^2}{8}}}{\sqrt{k^2+m^2}}dk\nonumber\\
    &=\frac{\sqrt{\pi}}{4\lambda_C^2}U\left(\frac32,2,\frac{\pi^3 R^2}{2\lambda_C^2}\right),
\end{align}
where
\begin{equation}
    U(a,b,z)=\frac{2}{\Gamma(a)}\int_0^\infty e^{-zt^2}t^{2a-1}(1+t^2)^{b-a-1}dt
\end{equation}
denotes \emph{ Tricomi's confluent hypergeometric function} \cite{gradshteyn2014table} and
\begin{equation}
    \lambda_C=\frac{2\pi}{m}
\end{equation}
is the Compton wavelength of a particle with mass $m$.

In a similar fashion by defining a Gaussian coherent amplitude distribution 
\begin{equation}
a(\kk)=\sqrt{\frac{|\kk|}{\omega(\kk)}}\frac{\exp(-\frac{2|\kk|^2}{\pi E^2})}{(\pi E/2)^{3/2}},
\end{equation}
with mean energy $E$ equal to the expectation value of the field Hamiltonian
\begin{equation}
    \hat{H}_\phi=\int \omega(\kk)\hat a^\dagger_{\kk}\hat a_{\kk}d^3\kk,
\end{equation} 
the mean value of the real part of Eq. (\ref{phi-a}) for a detector at rest at the origin of the coordinate system at a time $t_0=0$, depends on the the mean effective radius of the detector, the mean field energy and the mass of the field
\begin{align}\label{mass-re}
    \Re \langle\hat{\mathrm{a}}\rangle_a &=\sqrt{\frac{8}{\pi^4E^3}}\int_0^\infty{\frac{k^{5/2}}{\sqrt{k^2+m^2}}}\exp[-\frac{k^2}{2\sigma^2}]dk,\\\nonumber
    &=m\sqrt{\frac{2m^3}{\pi^4 E^3}}\Gamma\left(\frac74\right)U\left(\frac74,\frac94,\frac{m^2}{2\sigma^2}\right),
\end{align}
where for ease of notation we define
\begin{equation}
\frac{1}{\sigma^2}=\frac{4}{\pi E^2}+\frac{\pi R^2}{8}.
\end{equation}

In Fig. \ref{fig:1} (a)-(c) we present the $\ell_1$-cohering power of the field as a function of the detector's radius, and the mean energy of the field for various values of coupling constant. We observe that the ability of the field to generate coherence in an initially incoherent detector is reduced for increasing values of the coupling strength, and tends to zero in the asymptotic limit of a large detector radius and field energy. More importantly there are regions of the parameter space where the cohering power is identically equal to zero even in the non-asymptotic limit due to its oscillatory behavior.  Because of damping, these regions are hard to spot in the figure but are easily visible once the exponential factor in Eq. (\ref{coher:exp}) is removed such as in Fig. \ref{fig:1} (d) for example.
\begin{figure}
{\includegraphics[width=\linewidth]{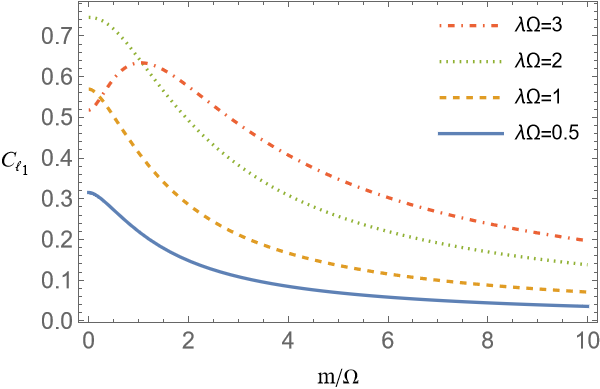}}
\caption{Dependence of $\ell_1$-cohering power of the field as a function of mass (measured with respect to the detector's energy gap $\Omega$) for a field in a coherent state with mean energy $E=\Omega$ and a detector with mean radius equal to $R=1/\Omega$.}
\label{mass:comp}
\end{figure}

From Eqs. (\ref{mass-com}) and (\ref{mass-re}) it is also evident that for a fixed coupling strength and a specific value of the detector's effective radius and energy of the field, a massive field's cohering power is damped less and is phase-shifted towards smaller values compared to the massless case. In Fig. \ref{mass:comp} we demonstrate the dependence of the cohering power on the field mass for different values of the coupling constant for a field with mean energy $E=\Omega$ and a detector with mean effective radius equal to $R=1/\Omega$. For couplings below $\lambda\Omega=2$ the cohering power of the field is in one-to-one correspondence with its mass.
%-------------------------------------------
%
%-------------------------------------------
\section{Decohering power of thermal fields}\label{SecV}
For a thermal field at an inverse temperature $\beta$
\begin{equation}
    \sigma_\phi=\frac{e^{-\beta\hat{H}_\phi}}{Z}
\end{equation}
with partition function $Z=\tr_\varphi e^{-\beta\hat{H}_\phi}$, let $\langle\hat X\rangle_\beta$ denote the dependence of the expectation value of field operator $\hat{X}$ on the temperature. Employing the same decomposition as in Eq. (\ref{decomposed field}) it can be shown that in this case
\begin{equation}\label{th-z}
    \langle e^{i2\lambda\hat\varphi_{f_0}}\rangle_\beta = e^{-2\lambda^2\langle\hat\varphi^2_{f_0}\rangle_\beta}.
\end{equation}
To see this we must first rewrite the left hand side following the same steps that led to Eq. (\ref{zBCH})
\begin{equation}\label{decomposed z}
    \langle e^{i2\lambda\hat\varphi_{f_0}}\rangle_\beta=e^{-2\lambda^2[\hat{\mathrm{a}},\hat{\mathrm{a}}^\dagger]}\langle e^{i2\lambda \hat{\mathrm{a}}^\dagger}e^{i2\lambda\hat{\mathrm{a}}}\rangle_\beta.
\end{equation}
To compute the expectation value on the right hand side we now Taylor expand $e^{i\lambda\hat{\mathrm{a}}}$ and $e^{i\lambda\hat{\mathrm{a}}^\dagger}$ to obtain
\begin{equation}
    \langle e^{i2\lambda \hat{\mathrm{a}}^\dagger}e^{i2\lambda\hat{\mathrm{a}}}\rangle_\beta=\\\sum_{m,m'=0}^\infty\frac{(i2\lambda)^{m+m'}}{(m!)(m'!)}\langle(\hat{\mathrm{a}}^\dagger)^m(\hat{\mathrm{a}})^{m'}\rangle_\beta.
\end{equation} 
Noting that because the field is diagonal in the energy basis we only need consider terms where $m=m'$ since all the rest will equal zero. We will now show that
\begin{equation}\label{identity}
    \langle(\hat{\mathrm{a}}^\dagger)^m(\hat{\mathrm{a}})^{m}\rangle_\beta = m!(\langle\hat{\mathrm{a}}^\dagger\hat{\mathrm{a}}\rangle_\beta)^m.
\end{equation}
With the help of the following identity
\begin{equation}
    e^{\hat{X}}\hat{Y}e^{-\hat{X}} = \hat{Y} +\frac{1}{2!}[\hat{X},\hat{Y}]+\frac{1}{3!}[\hat{X},[\hat{X},\hat{Y}]]+\cdots
\end{equation}
and the commutation relation between $\hat{a}_{\kk}$ and the field Hamiltonian
\begin{equation}
    [\hat{a}_{\kk},\hat{H}_\varphi]=\omega(\kk)\hat{a}_{\kk},
\end{equation}
we find that
\begin{equation}
    \hat{a}_\kk e^{-\beta\hat{H}_\varphi}=e^{-\beta \omega(\kk)}e^{-\beta\hat{H}_\varphi}\hat{a}_\kk.
\end{equation}
Using this and Eq. (\ref{commut:rel}) it is straightforward to show that
\begin{equation}
    \langle \hat{a}^\dagger_\kk \hat{a}_{\kk'}\rangle_\beta = \frac{e^{-\beta \omega(\kk')}}{1-e^{-\beta \omega(\kk')}}\delta(\kk-\kk'),
\end{equation}
which implies by induction
\begin{equation}\label{recursion}
    \langle\prod_{i=1}^m\hat{{a}}_{\kk_i}^\dagger\prod_{j=1}^m\hat{{a}}_{\kk'_j}\rangle_\beta =\sum_{i=1}^m\langle \hat{a}^\dagger_{\kk_{i}} \hat{a}_{\kk'_m}\rangle_\beta\langle\prod_{i\neq i'}\hat{{a}}_{\kk_i'}^\dagger\prod_{j=1}^{m-1}\hat{{a}}_{\kk'_j}\rangle_\beta.
\end{equation}
It follows that
\begin{equation}
    \langle(\hat{\mathrm{a}}^\dagger)^m(\hat{\mathrm{a}})^{m'}\rangle_\beta =m\langle\hat{\mathrm{a}}^\dagger\hat{\mathrm{a}}\rangle_\beta\langle(\hat{\mathrm{a}}^\dagger)^{m-1}(\hat{\mathrm{a}})^{m-1}\rangle_\beta.
\end{equation}
from which Eq. (\ref{identity}) can be obtained recursively. Finally
\begin{equation}
    \langle e^{i2\lambda \hat{\mathrm{a}}^\dagger}e^{i2\lambda\hat{\mathrm{a}}}\rangle_\beta =  e^{-4\lambda^2\langle\hat{\mathrm{a}}^\dagger\hat{\mathrm{a}}\rangle_\beta}=e^{2\lambda^2[\hat{\mathrm{a}},\hat{\mathrm{a}}^\dagger]}e^{-2\lambda^2\langle\hat{\varphi}^2_{f_0}\rangle_\beta}
\end{equation}
which completes the proof.

Looking back at Eq. (\ref{th-z}) and noting that $\mathcal{C}_{\ell_1}(\Phi)=\abs{\Im z}$ we observe, perhaps not surprisingly, that a thermal field is incapable of generating coherence through an instantaneous interaction. On the other hand its decohering power is equal to
\begin{equation}
    \mathcal{D}_{\ell_1}(\Phi)=1-e^{-\lambda^2I(\beta)}
\end{equation}
where
\begin{equation}
    I(\beta) = \frac{1}{(2\pi)^3}\int\frac{\abs{F(\kk)}^2}{\omega(\kk)}\coth\left( \frac{\beta\omega(\kk)}{2}\right)d^3\kk.
\end{equation}

\begin{figure}\begin{minipage}{\columnwidth}
\subfloat[]{\includegraphics[width=\textwidth]{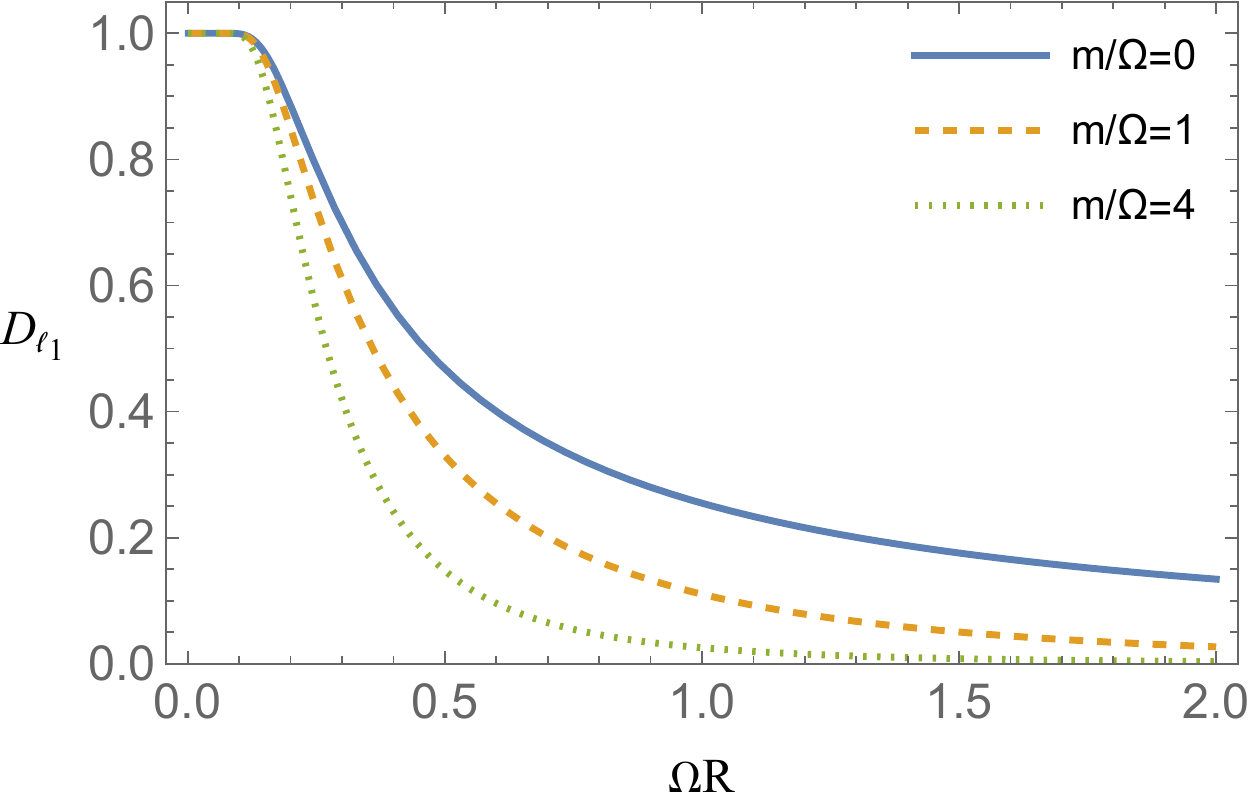}}\\
\subfloat[]{\includegraphics[width=\textwidth]{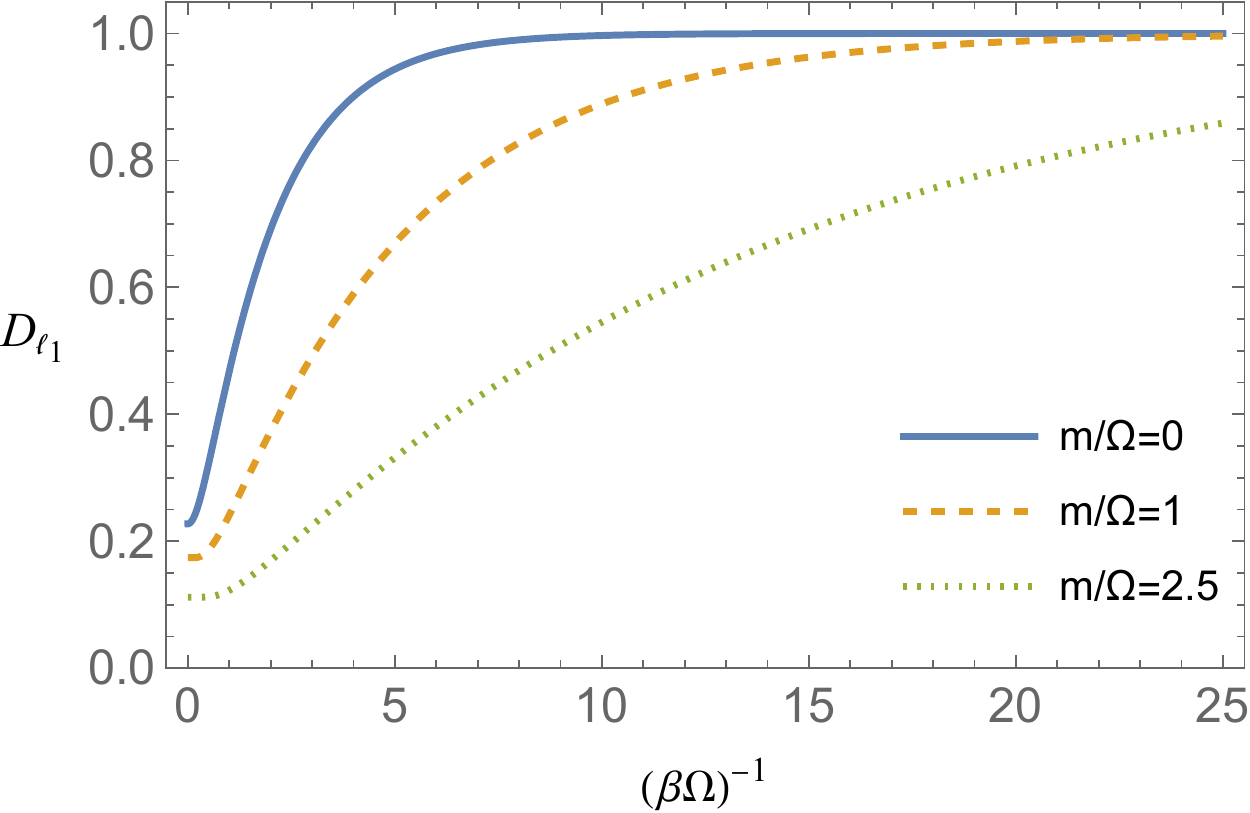}}
\end{minipage}
\caption{$\ell_1$-decohering power of a massive scalar field with respect to a) the radius of the detector for a field with inverse temperature $\beta=2/\Omega$ and a coupling constant $\lambda=\Omega$ and b) with respect to temperature for a detector with radius equal to $R=1/\Omega$ and a coupling constant $\lambda=2\Omega$.}\label{Fig:Thermal2}
\end{figure}

Since
\begin{equation}
    \frac{\partial{I(\beta)}}{\partial R}<0, \quad \frac{\partial{I(\beta)}}{\partial m}<0, \quad
    \frac{\partial{I(\beta)}}{\partial \beta}<0,
\end{equation}
the decohering power decreases for increasing values of the detector's radius and the mass of the field, while it increases with temperature. This is evident in Fig. \ref{Fig:Thermal2} (a) and (b) where we present the $\ell_1$-decohering power of a thermal field as a function of the detector's mean radius and the field's temperature for a detector with the same Gaussian smearing function as in Eq. (\ref{smear:gauss}).
\section{Discussion}
Employing an instantaneous interaction between a two-level UDW detector and a massive scalar field we investigate the ability of the field to generate or destroy coherence in the detector. This non-perturbative approach allows for an exact examination of the effects that different parameters (such as the strength of the coupling constant, the size of the detector, the energy of the field and its temperature for example) have on the cohering and decohering power of the induced quantum evolution channel.

In the case of coherence generation by a coherent field state it was shown that the success of the process depends on the size of the detector. More specifically, apart from the point-like $R<<R_0$, and macroscopic limit $R>>R_0$, where $R_0$ is some characteristic wavelength (e.g., the transition wavelength $\frac{2\pi}{\Omega}$ in the massless and Compton wavelength $\frac{2\pi}{m}$ in the massive case) there exist non-trivial values of the detector's radius for which it is impossible to generate any amount of coherence between its energy levels (Fig. \ref{fig:1} (d)). This phenomenon which manifests itself even in the case of a moderately weak coupling, as in Fig. \ref{fig:1} (a), demonstrates how the size of the system, which we wish to bring into a superposition of states, needs to be taken into consideration when designing experiments. It is expected that such effects will also be present in the generation of other quantum resources from the field, as in entanglement harvesting  \cite{VALENTINI1991321,Reznik2003,PhysRevA.71.042104,PhysRevD.92.064042}, for example. 

By calculating the decohering power of a thermal field, we investigated its effect on the set of maximally coherent states of the detector. In Fig. \ref{Fig:Thermal2}, it was demonstrated that for fixed values of the detector's radius, the field's temperature, and the coupling constant between the two, a massive field performs better than a massless one. Under a perturbative approach, similar results have also been reported in the limiting case of a point-like detector interacting weakly with the field in \cite{Huang2022}. 

For a suitable choice of detector radius, field energy and coupling strength it is possible to infer the mass of the field by measuring the amount of coherence present in the detector. This is evident from Fig. \ref{mass:comp}, where for a detector with $R=1/\Omega$, a field with energy $E=\Omega$ and coupling constants below $\lambda=2/\Omega$ the cohering power of the field is in one-to-one correspondence with its mass. The same conclusion can be made by studying the decohering power of a thermal field. Similar approaches have previously been employed for distinguishing the kinematic state of a detector interacting with a massive field by measuring its transition probability \cite{Quantaccel}, for determining the distance of closest approach for two accelerating detectors by studying the amount of entanglement that they can harvest from the field \cite{Salton_2015} and for probing the mass of an axion dark matter field by measuring the coherence stored in a detector \cite{coh:axions}.

The UDW Hamiltonian in Eq. \eqref{monopole} contains all of the essential features of the interaction of matter with an electromagnetic field \cite{Wavepacket:det,CHLI2}, where the analogue of a massive field in this case is a Proca field \cite{jackson1999classical}. Since massive electromagnetic theory is equivalent to Maxwell theory in Proca metamaterials \cite{Procameta}, the above results could find application for the construction of novel technologies such as new types of quantum memories communication channels and sensors. For this reason a more complete investigation of cohering and decohering effects, for detectors interacting continuously with massive fields, by making use of other non-perturbative methods \cite{bruschi2013,MartinezNonPert} would certainly be of interest.
%/////////////////////////////////////////////////////////////////////////////////////////////////////////
\acknowledgments{}
N. K. K. acknowledges the support and hospitality of the Kenneth S. Masters foundation during preparation of this manuscript. D. M. 's research has been co-financed by Greece and the  European Union (European Social Fund-ESF) through the Operational Programme ``Human Resources Development, Education and Lifelong Learning" in the context of the project ``Reinforcement of Postdoctoral Researchers - 2nd Cycle" (MIS-5033021), implemented by the State Scholarships Foundation (IKY).
\appendix
\section{Generalized cohering power}\label{first appendix}
Instead of the cohering power one could enquire whether it is possible to harness the coherence already present in a state to obtain a greater amount of coherence, by the action of a quantum operation, than what would be possible with only incoherent states. In this case one must specify the \emph{generalized cohering power} of the channel defined by
\begin{equation}
    \hat{\mathcal{C}}(\Phi) = \max_{\rho}[C(\Phi(\rho))-C(\rho)].
\end{equation}
It is obvious that $\mathcal{C}(\Phi)\leq\hat{\mathcal{C}}(\Phi)$. Depending on the choice of $C$ as a coherence measure, there exist channels such that the cohering power is strictly smaller than its generalized definition. For the $\ell_1$-norm of coherence in Eq. (\ref{l1-norm}) it was shown that for channels acting on qubits \cite{BuXiong}
\begin{equation}\label{generalized vs cohering}
    \mathcal{C}_{\ell_1}(\Phi) = \hat{\mathcal{C}}_{\ell_1}(\Phi),
\end{equation}
so this is in fact the maximum possible amount that can be obtained by the action of $\Phi$.
\bibliography{massive}
\end{document}